# On the Stability of the Endemic Equilibrium of A Discrete-Time Networked Epidemic Model[⋆]

**Fangzhou Liu** [∗] **Shaoxuan Cui** [∗] **Xianwei Li** [∗∗] **Martin Buss** [∗]

[∗] *Technical University of Munich, Munich, 80333 Germany (e-mail: {fangzhou.liu; shaoxuan.cui; mb}@tum.de).*
[∗∗] *Shanghai Jiaotong University, Shanghai, 200240 China (e-mail: lixianwei1985@gmail.com).*

**Abstract:** Networked epidemic models have been widely adopted to describe propagation phenomena. The endemic equilibrium of these models is of great significance in the field of viral marketing, innovation dissemination, and information diffusion. However, its stability conditions have not been fully explored. In this paper we study the stability of the endemic equilibrium of a networked Susceptible-Infected-Susceptible (SIS) epidemic model with heterogeneous transition rates in a discrete-time manner. We show that the endemic equilibrium, if it exists, is asymptotically stable for any nontrivial initial condition. Under mild assumptions on initial conditions, we further prove that during the spreading process there exists no overshoot with respect to the endemic equilibrium. Finally, we conduct numerical experiments on real-world networks to demonstrate our results.

## 1. INTRODUCTION

Originating from epidemiology, the study of epidemic spreading processes has become an attractive area due to its wide applications in propagation phenomena, such as computer virus dissemination and information diffusion Draief and Massouli (2010); Kephart and White (1991); Nowzari et al. (2016). In order to mathematically model such a process, multitudes of epidemic models incorporating different compartments have been proposed, e.g., susceptible-infected-susceptible (SIS), susceptible-infected-recovered-susceptible (SIRS), and susceptible-exposed-infected-susceptible (SEIR) Kermack and McKendrick (1927); Li et al. (2017). In addition, taking into consideration the local influence, networked epidemic models which emphasize individual dynamics have been introduced. In this case, epidemic diffusion process can be generally described as a graph-based Markov chain. Particularly, for the infection process of one agent, the transition rate is dependent on the states of her neighbors Liu and Buss (2016); Qu and Wang (2017). However, the exact Markov chain model faces difficulties due to the great number of states especially in large-scale networks. Thus in the seminal work Mieghem et al. (2009), a continuous-time $N$-intertwined SIS model has been proposed by implementing mean-field approximation (MFA). By reducing the number of states from $2^N$ to $N$, this model not only tackles the issue due to high dimensions but also exhibits acceptable accuracy in the sense of approximation. In Paré et al. (2018), a discrete-time version of the $N$-intertwined SIS model is introduced and further validated by real data from the fields of epidemic spreading and information diffusion. In this paper, we investigate this discrete-time networked SIS model and focus on the stability analysis of its nonzero equilibrium.

### 1.1 Related Work

Epidemic models in an SIS manner have been widely studied in both continuous-time and discrete-time Fall et al. (2007); Wang et al. (2003). The $N$-intertwined SIS model on an undirected network with homogeneous transition rates in Mieghem et al. (2009) is known to be the first probability-based networked epidemic model. It is then extended in Khanafer et al. (2014); Mieghem and Omic (2014) by taking into consideration directed communication topology and heterogeneous transition rates. Further extensions incorporate the SIS model on time-varying networks and bi-virus competing SIS models, where the stability conditions on *disease-free* (trivial) equilibrium and *endemic* (nontrivial) equilibrium are highlighted Liu et al. (2019); Paré et al. (2017).

A discrete-time networked SIS model with heterogeneous transition rates is introduced in Paré et al. (2018) by applying Euler method to the $N$-intertwined SIS model. The authors confirm the stability of the disease-free equilibrium under the condition regarding the transition rates and network connections. In addition, the endemic equilibrium is proved to be existed under certain conditions and possesses all positive components. However, the stability analysis on the endemic equilibrium has not been fully explored. Very recent work Prasse and Van Mieghem (2019) shows that

[⋆] This work was partially supported by the joint Sino-German project "Control and Optimization for Event-triggered Networked Autonomous Multi-agent Systems" funded by the German Research Foundation (DFG) and the National Science Foundation of China (NSFC).

the endemic equilibrium is exponentially stable but limited to special initial conditions. To the best our knowledge, there exist no results on the stability of the endemic equilibrium of the networked discrete-time SIS model in Paré et al. (2018) for general initial conditions.

## 1.2 Contribution

Our contribution is twofold. Firstly, we prove that the endemic equilibrium of the model in Paré et al. (2018) is asymptotically stable for any nonzero initial conditions. This fundamental result solves the open problem and completes the analysis of the model in Paré et al. (2018). Since the model in Paré et al. (2018) has been validated by real-world data, its detailed analysis could deepen the understanding of practical spreading processes. The study of the endemic equilibrium, as an essential part, may provide more insights in the field of viral marketing and information diffusion where nonzero steady-states are desired. Secondly, we consider two representative initial conditions, i.e., compared with the endemic equilibrium, all agents are in 1) lower infection level and 2) higher infection level. We show that during the spreading process there exists no overshoot with respect to the endemic equilibrium. Thus the results in Prasse and Van Mieghem (2018) for the model with homogeneous transition rates is extended to the model with heterogeneous transition rates. Based on this property, the domain of attraction in Prasse and Van Mieghem (2019) is enlarged.

The remainder of this paper is organized as follows: In Section 2, preliminaries on graph theory and nonnegative matrices are provided. We then introduce the networked discrete-time SIS model, as well as necessary properties and assumptions. The main results regarding the stability of the endemic equilibrium are presented in Section 3. Numerical experiments on real networks are conducted in Section 4.

*Notations:* Let $\mathbb{R}$ and $\mathbb{N}$ be the set of real numbers and nonnegative integers, respectively. Given a matrix $M \in \mathbb{R}^{n \times n}$, $\rho(M)$ is the spectral radius of $M$. For a matrix $M \in \mathbb{R}^{n \times r}$ and a vector $a \in \mathbb{R}^n$, $M_{ij}$ and $a_i$ denote the element in the $i$th row and $j$th column and the $i$th entry, respectively. For any two vectors $a, b \in \mathbb{R}^n$, $a \gg (\ll) b$ represents that $a_i > (<) b_i, i = 1, \ldots, n$; $a > (<) b$ means that $a_i \geq (\leq) b_i, i = 1, \ldots, n$ and $a \neq b$; and $a \geq (\leq) b$ means that $a_i \geq (\leq) b_i, i = 1, \ldots, n$ or $a = b$. These component-wise comparisons are also applicable for matrices with the same dimension. Vector $\mathbf{1}$ ($\mathbf{0}$) represents the column vector of all ones (zeros) with appropriate dimensions.

## 2. PRELIMINARIES AND MODEL DESCRIPTION

In this section, we briefly introduce the background knowledge of graph theory and properties of nonnegative matrices. Then the networked discrete-time SIS model and its properties are presented.

### 2.1 Graph Theory

Consider a weighted directed graph $\mathcal{G} = (\mathcal{V}, \mathcal{E}, A)$ where $\mathcal{V}$ and $\mathcal{E} \subseteq \mathcal{V} \times \mathcal{V}$ are the set of vertices and the set of edges, respectively. $A = [a_{ij}] \in \mathbb{R}^{N \times N}$ is the nonnegative adjacency matrix. $a_{ij} > 0$ if and only if there exists an edge from node $j$ to node $i$. In this paper, we confine ourselves that $\mathcal{G}$ is strongly connected. The adjacency matrix $A$ is irreducible if and only if the associated graph $\mathcal{G}$ is strongly connected.

### 2.2 Properties of Nonnegative Matrices

A matrix $M$ is nonnegative if all the entries of $M$ is nonnegative. We now introduce the Perron-Frobenius Theorem for irreducible nonnegative matrices.

*Lemma 1.* (Varga, 2000, Theorem 2.7) Given that a square matrix $M$ is an irreducible nonnegative matrix. The the following statements hold

i) $M$ has a positive real eigenvalue equal to its spectral radius $\rho(M)$.
ii) $\rho(M)$ is a simple eigenvalue of $M$.
iii) There exist a unique right eigenvector $x \gg 0$ and a unique left eigenvector $y^\top \gg 0$ corresponding to $\rho(M)$.

The following lemma collects the properties of nonnegative matrices which are necessary for this paper.

*Lemma 2.* (Horn and Johnson, 2013, Corollary 8.1.29 and 8.1.30) Given a nonnegative matrix $M \in \mathbb{R}^{n \times n}$, a positive vector $x \in \mathbb{R}^n$ ($x \gg \mathbf{0}$) and a nonnegative scalar $\mu$, there hold: if $\mu x \ll Mx$, then $\mu < \rho(M)$; if $\mu x \gg Mx$, then $\mu > \rho(M)$; and if $\mu x = Mx$, then $\mu = \rho(M)$.

### 2.3 Networked Discrete-Time SIS Model

In this paper, we consider the networked discrete-time SIS model in Paré et al. (2018) on an $N$-node network $\mathcal{G} = (\mathcal{V}, \mathcal{E}, A)$ as follows.

$$x_i(k+1) = x_i(k) + h\left((1 - x_i(k))\sum_{j=1}^N \beta_{ij} x_j(k) - \delta_i x_i(k)\right), \quad (1)$$

where $x_i(k)$ represents the infection probability of agent $i$ or the infection proportion of group $i$ at time instance $k$; $\beta_{ij} := \beta_i a_{ij}$; $\beta_i$ and $\delta_i$ are the infection rate and the curing rate of node $i$, respectively; and $h$ is the sampling period. The compact form of (1) reads

$$x(k+1) = [I - h((I - \text{diag}(x(k)))B - D)]x(k), \quad (2)$$

where $x(k) = [x_1(k), x_2(k), \ldots, x_N(k)]^\top$, $B = [\beta_{ij}]_{N \times N}$, and $D = \text{diag}(\delta_1, \delta_2, \ldots, \delta_N)$. Evidently, $\mathbf{0}$ is an equilibrium of the model in (2). It plays an significant role in epidemic spreading processes, since it represents the disease-free case. The non-trivial equilibrium ($x^* \neq \mathbf{0}$), on the contrary, is named the *endemic equilibrium*. The endemic equilibrium is of great importance in information diffusion and viral marketing where the nonzero steady-states are desired.

The following assumptions are assumed to hold throughout this paper.

*Assumption 1.* The initial condition satisfies $x(0) \in [0, 1]^N$.

*Assumption 2.* For every $i \in \mathcal{V}$, we have $\beta_i > 0$ and $\delta_i \geq 0$.

*Assumption 3.* The sampling period $h$ is positive and for every node $i \in \mathcal{V}$, there holds $h(\delta_i + \sum_{j=1}^N \beta_{ij}) \leq 1$.

Under these assumptions, we obtain the following lemma.

*Lemma 3.* Suppose that Assumptions 1, 2, and 3 hold. Given the networked discrete-time SIS model in (2), there holds $x(k) \in [0,1]^N$ for every $k \in \mathbb{N}$.

**Proof.** We prove Lemma 3 by induction. If $k=0$, we have $x(0) \in [0,1]^N$ by Assumption 1. Suppose $x(k) \in [0,1]^N$ holds for some $k > 0$. By using the dynamics in (1), Assumptions 2 and 3, we have

$$x_i(k+1) \leq x_i(k)(1-h\delta_i) + (1-x_i(k))\left(h\sum_{j=1}^N \beta_{ij}\right)$$

$$\leq 1 - h\delta_i + h\sum_{j=1}^N \beta_{ij} \leq 1$$

and
$$x_i(k+1) \geq x_i(k)(1-h\delta_i) \geq 0.$$

Thus we conclude that $x(k) \in [0,1]^N$ for every $k \in \mathbb{N}$.

Lemma 3 implies that $[0,1]^N$ is a positive invariant set of $x(k)$. Bearing in mind that $x_i(k)$ is the infection probability or the infection proportion, it is natural to limit the value within the interval $[0,1]$. In this sense, the model in (2) is well-defined. Note that in order to obtain Lemma 3, Assumption 3 can be relaxed to the assumption that for every node $i \in \mathcal{V}$, there hold $h\delta_i \leq 1$ and $h\sum_{j=1}^N \beta_{ij} \leq 1$. This is adopted in Paré et al. (2018) to build a generic model. Compared with this assumption, Assumption 3 is also reasonable and applicable when the sampling period $h$ is short. In this paper, we accept Assumption 3 because it does lead to a well-defined model and most of our results rely on this technical setting.

Further properties of the equilibria of the model in (2) are concluded in the following lemma.

*Lemma 4.* (Paré et al., 2018) Under Assumptions 1, 2, and 3, the following statements hold for the dynamics in (2):

 i) If $\rho(I - hD + hB) \leq 1$, then $\mathbf{0}$ is the only equilibrium of the model in (2) and asymptotically stable with domain of attraction $[0,1]^N$.
 ii) If $\rho(I - hD + hB) > 1$, then the model in (2) possesses two equilibria, $\mathbf{0}$ and $x^*$. Moreover, there holds $x^* \gg \mathbf{0}$.

Lemma 4 provides the condition for the existence of an endemic equilibrium $x^*$ with all strictly positive components. However, It is still an open problem whether the endemic equilibrium is stable or not. In the remainder of this paper, we solve this problem and provide further properties of the network discrete-time SIS model. To avoid ambiguity, we discuss the endemic equilibrium only when it exists, i.e., the following assumption holds

*Assumption 4.* For the dynamics (2), we have $\rho(I - hD + hB) > 1$.

## 3. MAIN RESULTS

Before embarking on the stability analysis, we present further properties of the endemic equilibrium. The boundaries of the endemic equilibrium are provided in the following lemma.

*Lemma 5.* Suppose that Assumptions 1-4 hold. Then the steady-state $x_i^*$ of any node $i \in \mathcal{V}$ satisfies

$$1 - \frac{\delta_m}{\sum_{j=1}^N \beta_{mj}} \leq x_i^* \leq 1 - \frac{\delta_i}{\delta_i + \sum_{j=1}^N \beta_{ij}}, \quad (3)$$

where $m = \arg\min_{i \in \mathcal{V}}\{x_i^*\}$.

**Proof.** By substituting the endemic equilibrium $x_i^*$ into the model (1), we obtain

$$(1-x_i^*)\sum_{j=1}^N \beta_{ij} x_j^* = \delta_i x_i^*. \quad (4)$$

It follows that

$$x_i^* = 1 - \frac{\delta_i}{\delta_i + \sum_{j=1}^N \beta_{ij} x_j^*}. \quad (5)$$

In light of $x_i^* \leq 1, \forall i \in \mathcal{V}$, it is clear that

$$x_i^* \leq 1 - \frac{\delta_i}{\delta_i + \sum_{j=1}^N \beta_{ij}}. \quad (6)$$

Let $m = \arg\min_{i \in \mathcal{V}}\{x_i^*\}$. Thus by (5), we have

$$x_m^* \geq 1 - \frac{\delta_m}{\delta_m + x_m^* \sum_{j=1}^N \beta_{mj}} = \frac{x_m^* \sum_{j=1}^N \beta_{mj}}{\delta_m + x_m^* \sum_{j=1}^N \beta_{mj}}. \quad (7)$$

Bearing in mind that $x_i^* > 0, \forall i \in \mathcal{V}$ by Lemma 4, it yields that

$$x_m^* \geq 1 - \frac{\delta_m}{\sum_{j=1}^N \beta_{mj}}. \quad (8)$$

By combining (6) and (8), it is evident that the relation (3) holds.

In Lemma 5, we obtain explicit boundaries of the endemic equilibrium compared with Lemma 4 ($x^* \gg \mathbf{0}$), which is necessary for our further analysis. Equipped with Lemma 5, we present the following results on the stability of the endemic equilibrium of the discrete-time networked SIS model in (2).

*3.1 Stability of the Endemic Equilibrium for Special Initial Conditions*

Inspired by Prasse and Van Mieghem (2019), we start the stability analysis of the endemic equilibrium from special cases. Specifically, we consider two categories of representative initial conditions: compared with the endemic equilibrium, all agents are in 1) lower infection level and 2) higher infection level, i.e., the initial conditions are in the following two respective sets

$$\begin{aligned}\mathcal{D}_l &= \{x(0) \in [0,1]^N : \mathbf{0} < x(0) \leq x^*\} \\ \mathcal{D}_h &= \{x(0) \in [0,1]^N : x^* < x(0) \leq \mathbf{1}\}.\end{aligned} \quad (9)$$

These initial conditions commonly exist in real-world spreading processes. Specifically, initial conditions in $\mathcal{D}_l$ appear at early stage in the epidemic or information diffusion processes when all the nodes are in a low infection level. The initial conditions in $\mathcal{D}_h$, however, mimic the scenario when viruses have been widely spread and we should take actions to reduce the infection level. Note that $\mathbf{0}$ is excluded from $\mathcal{D}_l$, due to the fact that $x(k)$ will stay at the origin if $x(0) = \mathbf{0}$.

The following proposition confirms the stability of the endemic equilibrium given the two aforementioned initial conditions.

*Proposition 1.* Suppose that Assumptions 1-4 hold. Given the networked discrete-time SIS model in (2) with endemic equilibrium $x^*$, the endemic equilibrium is exponentially stable with domain of attraction $\mathcal{D} = \mathcal{D}_l \cup \mathcal{D}_h$. Moreover, the following two statements hold

i) if $x(0) \in \mathcal{D}_l$, then $x_i(k) \leq x_i^*, \forall i \in \mathcal{V}, k \in \mathbb{N}$,
ii) if $x(0) \in \mathcal{D}_h$, then $x_i(k) \geq x_i^*, \forall i \in \mathcal{V}, k \in \mathbb{N}$.

**Proof.** The proof for the case when $x(0) \in \mathcal{D}_l$ has been detailed in Prasse and Van Mieghem (2019) and we save it for triviality. Now we focus on the case when $x(0) \in \mathcal{D}_h$. Since $x_i^*$ is an equilibrium of (1), there holds

$$x_i^* = x_i^* + h\left((1-x_i^*)\sum_{j=1}^N \beta_{ij} x_j^* - \delta_i x_i^*\right)$$

In conjugation with Assumption 2 and Lemma 4, we have $x_i^* < 1, \forall i \in \mathcal{V}$. It follows that

$$\sum_{j=1}^N \beta_{ij} x_j^* = \frac{\delta_i x_i^*}{1 - x_i^*}. \tag{10}$$

Let $z_i(k) = x_i(k) - x_i^*$. By using (10), the error dynamics reads

$$z_i(k+1) = \left(1 - h\delta_i - h\sum_{j=1}^N \beta_{ij} x_j(k)\right) z_i(k) \\ + (1-x_i^*)h\sum_{j=1}^N \beta_{ij} z_j(k). \tag{11}$$

We first prove statement ii) is true based on dynamics (11) by induction. If $k = 0$, it straightforward that $x_i(0) \geq x_i^*$ since $x(0) \in \mathcal{D}_h$. Suppose for some $k > 0$, there holds $x(k) \in \mathcal{D}_h$. It implies that $z_i(k) \geq 0, \forall i \in \mathcal{V}$. Taking into account Assumption 3 and Lemma 4, it implies that the two summands in (11) are both nonnegative. It yields that $z_i(k) \geq 0, \forall i \in \mathcal{V}$. This is equivalent to statement ii).

It remains to show $x^*$ is exponentially stable if $x(0) \in \mathcal{D}_h$. Equivalently, we prove that the system (11) is exponentially stable. By using the relation (10), we can rewrite (11) as

$$z_i(k+1) = \left(1 - \frac{h\delta_i}{1-x_i^*}\right) z_j(k) + (1-x_i^*-z_i(k))h\sum_{j=1}^N \beta_{ij} z_j(k). \tag{12}$$

Denote $z(k) = [z_1(k), z_2(k), \ldots, z_N(k)]^\top$. The matrix form of (12) reads

$$z(k+1) = \Xi z(k) - h\,\mathsf{diag}(z(k))Bz(k), \tag{13}$$

where

$$\Xi = I - \mathsf{diag}\left(\frac{h\delta_1}{1-x_1^*}, \frac{h\delta_2}{1-x_2^*}, \ldots, \frac{h\delta_N}{1-x_N^*}\right) \\ + h\,\mathsf{diag}(\mathbf{1}-x^*)B. \tag{14}$$

We then show the exponential stability by comparison principle. In accordance to statement ii) and (13), there holds

$$0 \leq z(k+1)_i \leq [\Xi z(k)]_i, \forall i \in \mathcal{V}. \tag{15}$$

Thus there exists a dynamical system $\bar{z}(k+1) = \Xi \bar{z}(k)$, with $\bar{z}(0) = z(0)$, such that $\bar{z}_i(k) \geq \bar{z}_i(0), \forall i \in \mathcal{V}, k \in \mathbb{N}$.

It follows that we only need to prove $\rho(\Xi) < 1$. For every $i \in \mathcal{V}$, we have

$$\Xi_{ii} = 1 + \frac{h\delta_i}{x_i^* - 1} + (1-x_i^*)h\beta_{ii} \\ \geq \frac{1 - h\delta_i - x_i^*}{1 - x_i^*}.$$

In light of Assumption 3, Lemma 5, and the relation (10), we attain

$$\Xi_{ii} \geq \frac{1 - h\delta_i/(h\delta_i + h\sum_{j=1}^N \beta_{ij}) - x_i^*}{1 - x_i^*} \geq 0.$$

Notice that

$$\Xi_{ij} = h(1-x_i^*)\beta_{ij} \geq 0, \forall i, j \in \mathcal{V}, i \neq j.$$

It yields that $\Xi_{ij} \geq 0, \forall i, j \in \mathcal{V}$. Bearing in mind that $\mathcal{G}$ is irreducible and Assumption 3 holds, it implies that $\Xi$ is a nonnegative irreducible matrix. We then calculate the $i$th entry of $(\Xi - I)x^*$ as follows

$$[(\Xi - I)x^*]_i = -\frac{h\delta_i x_i^*}{1-x_i^*} + h(1-x_i^*)\sum_{j=1}^N \beta_{ij} x_j^*.$$

According to the relation (10) and Lemma 4, we have

$$[(\Xi - I)x^*]_i = -\frac{h\delta_i x_i^*}{1-x_i^*} + h\delta_i x_i^* < 0.$$

It yields that $\Xi x^* \ll x^*$. By Lemma 2, it implies that $\rho(\Xi) < 1$. Therefore the trajectories of $\bar{z}$ converge to the origin exponentially. By comparison principle, the system in (11) is exponentially stable. This completes our proof.

*Remark 1.* Proposition 1 incorporates two fundamental results under the initial condition $x(0) \in \mathcal{D}_l \cup \mathcal{D}_h$. Firstly, the endemic equilibrium, if it exists, is exponentially stable with domain of attraction $\mathcal{D}$. Secondly, there exists no overshoot with respect to the endemic equilibrium. However, it does not lead to the monotonicity of the trajectory of $x(k)$, which will be illustrated in Section 4 in detail. Note that the results under the initial condition $x(0) \in \mathcal{D}_l$ has been reported in Prasse and Van Mieghem (2019). Statement ii) in Proposition 1 has been obtained in Prasse and Van Mieghem (2018) for dynamics with homogeneous transition rates. Here we not only extend the results to heterogeneous epidemic models but also prove the exponential stability under initial condition $x(0) \in \mathcal{D}_h$. Based on the analysis on special initial conditions, we then generalize the results to any nontrivial initial condition.

*Remark 2.* In Ahn and Hassibi (2013), the authors show that the endemic equilibrium is not stable by providing a counterexample where the system converges to a cycle. This result, however, is not applicable to the dynamics in of (13) because Assumption 3 is negated. Moreover, under their configurations, $[0,1]^N$ is no longer a positive invariant set which is inappropriate from practical perspective.

*3.2 Global Stability*

Until now, we only consider the stability of the endemic equilibrium of the model (13) for certain initial conditions. It is of great importance to study the stability in the set $[0,1]^N$ from both theoretical and practical perspectives. Thus we provide the following theorem.

*Theorem 1.* Suppose that Assumptions 1-4 hold. Given the networked discrete-time SIS model on graph $\mathcal{G}$, the endemic equilibrium is asymptotically stable with domain of attraction $[0,1]^N \setminus \{\mathbf{0}\}$ and the disease free equilibrium is stable if and only if $x(0) = \mathbf{0}$.

**Proof.** Let us rewrite the dynamical system (13) in the following form.
$$z(k+1) = \Phi(k)z(k), \qquad (16)$$
where
$$\Phi(k) = I - \mathsf{diag}\left(\frac{h\delta_1}{1-x_1^*}, \frac{h\delta_2}{1-x_2^*}, ..., \frac{h\delta_N}{1-x_N^*}\right) \qquad (17)$$
$$+ \mathsf{diag}(\mathbf{1} - x(k))hB$$
Construct a square matrix $F$ as
$$F = I - \mathsf{diag}\left(\frac{h\delta_1}{1-x_1^*}, \frac{h\delta_2}{1-x_2^*}, ..., \frac{h\delta_N}{1-x_N^*}\right) + hB. \qquad (18)$$
Evidently, we have $\Phi(k) \leq F$ and equality is valid if and only if $x(k) = \mathbf{0}$. By using the relation (10), matrix $F$ can be rewritten as
$$F = \begin{bmatrix} 1 - \sum_{j \neq 1} h\beta_{1j}\frac{x_j^*}{x_1^*} & h\beta_{12} & \cdots & h\beta_{1N} \\ h\beta_{21} & 1 - \sum_{j \neq 2} h\beta_{ij}\frac{x_j^*}{x_2^*} & \cdots & h\beta_{2N} \\ \vdots & \vdots & \ddots & \vdots \\ h\beta_{N1} & h\beta_{N2} & \cdots & 1 - \sum_{j \neq N} h\beta_{ij}\frac{x_j^*}{x_N^*} \end{bmatrix}$$
Let $\mu = [1, \frac{x_2^*}{x_1^*}, \ldots, \frac{x_N^*}{x_1^*}]^\top$. Notice that
$$F\mu = \mu. \qquad (19)$$
Since $F$ is an irreducible nonnegative matrix and $\mu \gg \mathbf{0}$, there holds $\rho(F) = 1$ by Lemma 2. By Lemma 1, $F$ possesses a positive left eigenvector $v^\top$ corresponding to its spectral radius, i.e., $v^\top F = v^\top$. Then we discuss the following two cases.

*Case 1:* If $x(0) = \mathbf{0}$, by using the dynamics (2), it follows that $x(k) = \mathbf{0}, \forall k \in \mathbb{N}$. Thus if $x(0) = \mathbf{0}$, the trajectories of $x(k)$ stay at the origin and do not converge to the endemic equilibrium. In addition, if $x(k) = \mathbf{0}$ and $k \geq 1$, by the dynamical system (1), we have
$$(1-h\delta_i)x_i(k-1) + h(1-x_i(k-1))\sum_{j=1}^{N}\beta_{ij}x_j(k) = \mathbf{0}. \qquad (20)$$
It is clear that both of the summands are nonnegative. In light of Assumptions 2 and 3, it follows that $x(k-1) = \mathbf{0}$. Thus $x(m) = \mathbf{0}, \forall m \in \mathbb{N}$, if there exists certain $k$ such that $x(k) = \mathbf{0}$. Therefore, $\mathbf{0}$ is stable if and only if $x(0) = \mathbf{0}$.

*Case 2:* If $x(0) \neq \mathbf{0}$, we first show that there must exists a time instance $s$ such that $x(s) \gg \mathbf{0}$. Since $\mathcal{G}$ is strongly connected, $\sum_{j=1}^{N}\beta_{ij} > 0, \forall i \in \mathcal{V}$ by Assumption 2. It follows that $1 - h\delta_i > 0$ by Assumption 3. Let $x_l(0) > 0$ for certain $l \in \mathcal{V}$. In light of the dynamical system (1), we have $x_l(k) > 0, \forall k \in \mathbb{N}$. We then consider the worst case. Without loss of generality, we suppose only $x_p(0) > 0, p \in \mathcal{V}$ and other entries of $x(0)$ are 0. Based on the aforementioned derivation, $x_p(k)$ stays positive for any $k \in \mathbb{N}$. Since $\mathcal{G}$ is strongly connected, we can always find a node $q$ such that there exists an edge from $q$ to $p$, i.e., $a_{pq} > 0$. By utilizing the dynamical system (1), it is apparent that $x_q(1) = h\beta_{pq}x_q(0) > 0$. Similarly, all other entries turn to be positive in finite time steps thanks to the strong connectivity of the $N$-node graph $\mathcal{G}$.

Construct the following auxiliary system for any $k \geq s$.
$$y(k+1) = \Phi(k)y(k), \qquad (21)$$
with initial condition $y_i(s) = |z_i(s)|, \forall i \in \mathcal{V}$. Since $\Phi(k)$ is nonnegative, it is straightforward that $-y(k) \leq z(k) \leq y(k), \forall k \in \mathbb{N}$. Thus $z(k)$ is asymptotically stable, if the origin of the system (21) is asymptotically stable. Consider the Lyapunov function $V(k) = v^\top y(k)$. We can obtain its increment as follows.
$$\Delta V(k) = V(k+1) - V(k) = v^\top(\Phi(k) - I)y$$
$$= v^\top(\Phi(k) - F)y = -hv^\top \mathsf{diag}(x(k))By \qquad (22)$$
$$\leq 0.$$
Since $x, v \gg \mathbf{0}$ and the sum of each row in $B$ is positive, it implies that $\Delta V = 0$ if and only if $y = \mathbf{0}$. Thus by LaSalle's invariance principle, the origin of the model (21) is asymptotically stable. Based on comparison principle, we conclude that the model (16) is asymptotically stable for any non-zero initial condition. This is equivalent to the statement in Theorem 1.

Theorem 1 confirms that the endemic equilibrium of the system (2), if it exists, is asymptotically stable for any nontrivial initial condition. Thus the open problem on the stability of the endemic equilibrium is solved. Although mild assumptions are adopted, this theorem is applicable for most of the cases where the sampling period is short. Stability analysis based on relaxed assumptions will be investigated in the future work.

## 4. NUMERICAL EXAMPLES

In this section, we demonstrate our main results by numerical experiments. The simulations are conducted on a real-world network in Coleman (1964) with slight modification. This network describes the friendships between boys in a small highschool in Illinois. The largest strongly connected subgraph containing 67 nodes is utilized. For the convenience of simulation, we normalize the weights according to the number of in-neighbors, i.e., $\sum_{j=1}^{N} a_{ij} = 1, \forall i \in \mathcal{V}$. We take into account two types of parameters listed in the following table.

Table 1. Parameters in Model (1)

|  | $\beta_i$ | $\delta_i$ | $h$ |
| --- | --- | --- | --- |
| Parameters I | $(0.15, 0.25)$ | $(0.25, 0.35)$ | 1 |
| Parameters II | $(0.45, 0.55)$ | $(0.25, 0.35)$ | 1 |

The infection rates and curing rates are randomly selected from the corresponding intervals such that we end up with an SIS model with heterogeneous transition rates. Note that we have $\rho(I - hD + hB) < 1$ if Parameters I are adopted, while $\rho(I - hD + hB) > 1$ if Parameters II are utilized. In addition, Assumption 3 is satisfied. By using the same initial conditions $x(0) \in [0, 0.2]^N$ in the model (1) with Parameters I and II, we obtain the trajectories of $x(k)$ which are shown in Figures 1 and 2, respectively. Thus the results are consistent with Lemma 4 and confirm the existence of the endemic equilibrium if $\rho(I - hD + hB) > 1$. Furthermore, it is clear that the endemic equilibrium is stable when using Parameters II.

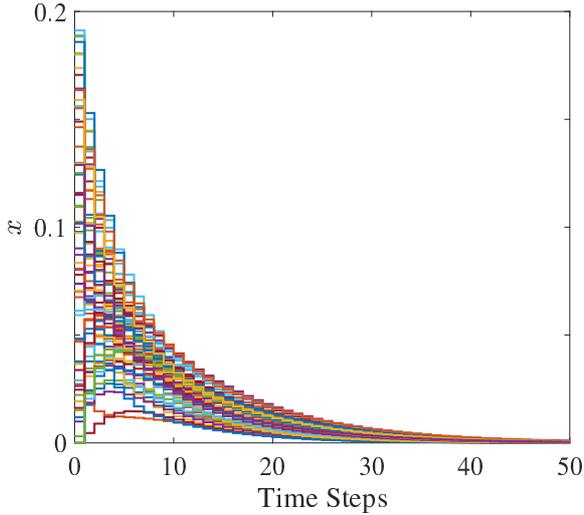

Fig. 1. The model in (2) with Parameters I and initial condition $x(0) \in [0, 0.2]^N$ converges to the disease-free equilibrium.

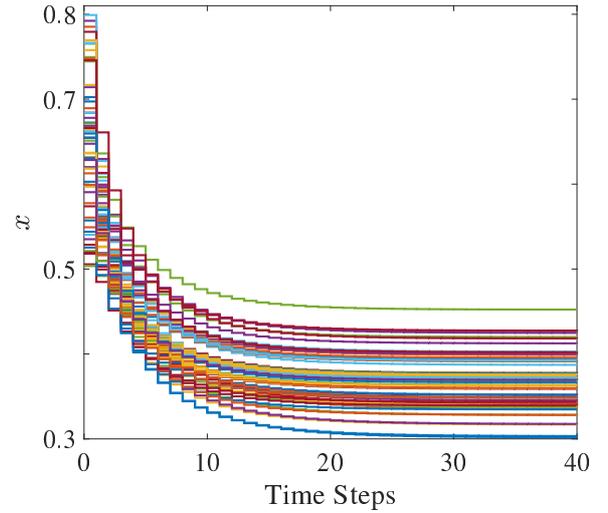

Fig. 3. The model in (2) with Parameters II and initial condition $x(0) \in (0.5, 0.8)^N$ converges to the endemic equilibrium. There exist no overshoots in the case $x_i(0) \geq x_i^*, \forall i \in \mathcal{V}$.

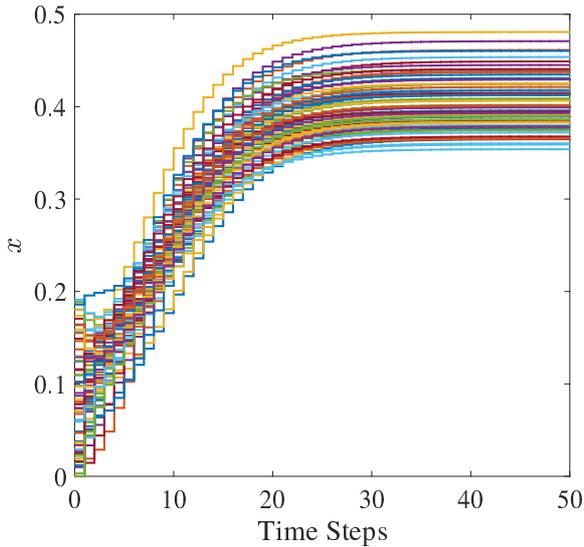

Fig. 2. The model in (2) with Parameters II and initial condition $x(0) \in [0, 0.2]^N$ converges to the endemic equilibrium. There exist no overshoots in the case $x_i(0) \leq x_i^*, \forall i \in \mathcal{V}$.

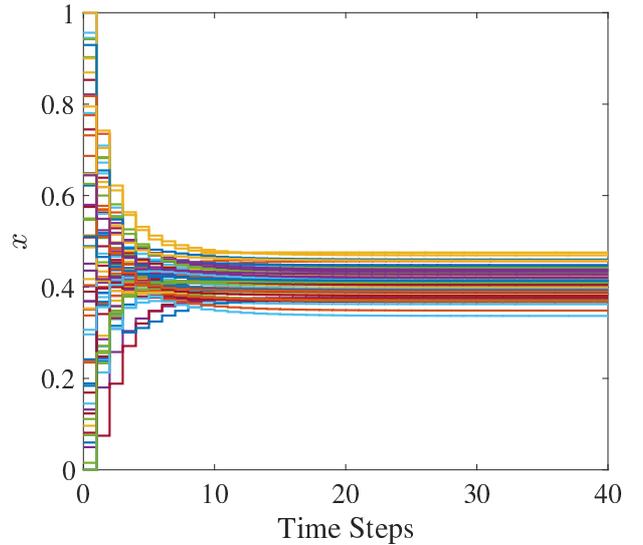

Fig. 4. The model in (2) with Parameters II and initial condition $x(0) \in [0, 1]^N$ converges to the endemic equilibrium.

Apart from the stability of the endemic equilibrium, Figure 2 also manifests that, there exists no overshoot with respect to the endemic equilibrium if the initial infection probabilities are small. In addition, we can observe that not all the trajectories are monotonically increasing. As a comparison, we conduct the simulation with large initial infection probabilities ($p(0) \in (0.5, 0.8)^N$) in model (1) with Parameters II. As is presented in Figure 3, $x(k)$ converges to the endemic equilibrium and $x_i(k) \geq x_i^*, \forall i \in \mathcal{V}, k \in \mathbb{N}$. These results support the two statements in Proposition 1. We further conduct the simulation with random initial condition within the set $[0, 1]^N$. Apparently, the trajectories in Figure 4 converge to the endemic equilibrium, which demonstrate the statement in Theorem 1.

## 5. CONCLUSION

In this paper we investigate the networked discrete-time SIS model in Paré et al. (2018). The main focus is the stability of the endemic equilibrium. We solve this open problem by providing rigorous proof and we confirm that the endemic equilibrium, if it exists, is asymptotically stable for any nontrivial initial condition under mild assumptions on the parameters. Additionally, for two kinds of initial conditions, i.e., high initial infection level and low initial infection level, we show that no overshoots with respect to the endemic equilibrium occur in the trajectories of all individual states.

Future work focuses on relaxing Assumption 3 such that the stability of the endemic equilibrium can be analyzed in more general situations.

REFERENCES

Ahn, H.J. and Hassibi, B. (2013). Global dynamics of epidemic spread over complex networks. In *the 52nd IEEE Conference on Decision and Control (CDC)*, 4579–4585.

Coleman, J.S. (1964). Introduction to mathematical sociology. *London Free Press Glencoe*.

Draief, M. and Massouli, L. (2010). *Epidemics and Rumours in Complex Networks*. Cambridge University Press, Cambridge.

Fall, A., Iggidr, A., Sallet, G., and Tewa, J.J. (2007). Epidemiological models and lyapunov functions. *Mathematical Modelling of Natural Phenomena*, 2(1), 55–73.

Horn, R.A. and Johnson, C.R. (2013). *Matrix Analysis*. Cambridge University Press.

Kephart, J. and White, S. (1991). Directed-graph epidemiological models of computer viruses. In *1991 IEEE Computer Society Symposium on Research in Security and Privacy*. Oakland, CA, USA.

Kermack, W.O. and McKendrick, A.G. (1927). A contribution to the mathematical theory of epidemics. *Proc. Roy. Soc. A*, 115(772), 700–721.

Khanafer, A., Basar, T., and Gharesifard, B. (2014). Stability properties of infected networks with low curing rates. In *American Control Conference*, 3579–3584.

Li, M., Wang, X., Gao, K., and Zhang, S. (2017). A survey on information diffusion in online social networks: Models and methods. *Information*, 8(118).

Liu, F. and Buss, M. (2016). Node-based sirs model on heterogeneous networks: Analysis and control. In *American Control Conference*, 2852–2857.

Liu, J., Paré, P.E., Nedić, A., Tang, C.Y., Beck, C.L., and Baśar, T. (2019). Analysis and control of a continuous-time bi-virus model. *IEEE Transactions on Automatic Control*. doi:10.1109/TAC.2019.2898515.

Mieghem, P.V. and Omic, J. (2014). In-homogeneous virus spread in networks. *arXiv preprint arXiv:1306.2588*.

Mieghem, P.V., Omic, J., and Kooij, R. (2009). Virus spread in networks. *IEEE/ACM Transactions on Networking*, 17(1), 1–14.

Nowzari, C., Preciado, V.M., and Pappas, G.J. (2016). Analysis and control of epidemics: A survey of spreading processes on complex networks. *IEEE Control Systems*, 36(1), 26–46.

Paré, P.E., Beck, C.L., and Nedić, A. (2017). Epidemic processes over time–varying networks. *IEEE Transactions on Control of Network Systems*, 5(3), 1322–1334.

Paré, P.E., Liu, J., Beck, C.L., Kirwan, B.E., and Basar, T. (2018). Analysis, estimation, and validation of discrete-time epidemic processes. *IEEE Transactions on Control Systems Technology*. doi:10.1109/TCST.2018.2869369.

Prasse, B. and Van Mieghem, P. (2018). Network reconstruction and prediction of epidemic outbreaks for nimfa processes. *arXiv preprint arXiv:1811.06741*.

Prasse, B. and Van Mieghem, P. (2019). The viral state dynamics of the discrete-time nimfa epidemic model. *arXiv preprint arXiv:1903.08027*.

Qu, B. and Wang, H. (2017). Sis epidemic spreading with heterogeneous infection rates. *IEEE Transactions on Network Science and Engineering*, 4(3), 177–186.

Varga, R. (2000). *Matrix Iterative Analysis*. Springer-Verlag.

Wang, Y., Chakrabarti, D., Wang, C., and Faloutsos, C. (2003). Epidemic spreading in real networks: An eigenvalue viewpoint. In *the 22nd International Symposium on Reliable Distributed Systems*, 25–34.